\documentclass[twocolumn]{revtex4}
\usepackage{graphicx}
\usepackage{epstopdf}
\usepackage{amsmath}
\usepackage{color}

\usepackage{gensymb}
\usepackage{setspace}
\usepackage{autobreak}
\usepackage[colorlinks,urlcolor=blue,citecolor=blue]{hyperref}

\begin{document}
	
		\title{Origin of Ferroelectricity and Superconductivity with Nontrivial Electronic Topology in Fluorinated Nb$_2$N}
		\author{Xin-Zhu Yin$^1$, Na Jiao$^{1,*}$, Jinlian Lu$^{3}$, Meng-Meng Zheng$^{1}$, Hong-Yan Lu$^{1,*}$, and Ping Zhang$^{1,2}$}
		\affiliation{
			$^1$School of Physics and Physical Engineering, Qufu Normal University, Qufu 273165, China \\
			$^2$Institute of Applied Physics and Computational Mathematics, Beijing 100088, China \\
			$^3$Department of Physics, Yancheng Institute of Technology, Yancheng, Jiangsu 224051, China
		}
	
	\begin{abstract}
		 
		Two-dimensional (2D) intrinsic superconductors with nontrivial topological band and vertical ferroelectricity exhibit fascinating characteristics to achieving electrostatic control of quantum phases. While, only a few such 2D materials have been theoretically predicted. In this work, based on first-principles calculations, we explore the superconductivity and ferroelectric properties in fluorinated 2D Nb$_2$N. In the stable Nb$_2$NF$_2$, H$_3$-Nb$_2$NF$_2$ breaks the spatial inversion symmetry, exhibiting vertical ferroelectric. More interestingly, it not only possesses intrinsic superconductivity with superconducting transition temperatures ($T_{c}$) of 10 K, but also exhibits nontrivial band topology. While, H$_1$-Nb$_2$NF$_2$ shows topological band and superconductivity with $T_{c}$ of 32 K, surpassing most of 2D conventional topological superconductors' candidates. Our research has enriched 2D superconducting materials with nontrivial band topology and ferroelectric properties, and provided a theoretical basis for the preparation of devices switching between superconducting and ferroelectric states with external electric field.
		
	\end{abstract}
	\maketitle
	\section{Introduction}
	
	Ferroelectric materials have great technological importance for realizing nonvolatile random-access memory. 2D ferroelectric materials have rich application field, including but not limited to 2D ferroelectric-based energy harvesters \cite{energy-harvesters}, tunnel junctions \cite{tunnel-junctions}, field-effect transistors \cite{transistors}, photodetectors \cite{photodetectors} and photocatalytic elements \cite{photocatalytic}. 2D ferroelectric metal not only possess a spontaneous and switchable electric polarization through the application of an external electronic field, but also exhibit metallic property. Recently, the advantages of 2D ferroelectric metal have been experimentally confirmed and theoretically predicted in 2D layered and van der Waals materials \cite{2019-MoTe2,2023-MoTe2}. Experimentally, the 2D topological semimetal WTe$_{2}$ was confirmed to exhibit spontaneous out-of-plane electric polarization, which can be switched using gate electrodes \cite{2018nature}. Several 2D materials that can sustain ferroelectric metal states have been theoretically predicted, such as monolayer CrN \cite{CrN}, two-unit-cell thick LiOsO$_3$ thin film \cite{LiOsO3}, 1T'-WTe$_2$ multilayer \cite{WTe2-1,WTe2-2}, $etc$. Thus, 2D ferroelectric metal offers compelling advantages in tunable electronic behavior via conventional electrostatic techniques to exploit ferroelectricity to control other electronic states.
	 
    A tunable superconductivity with ferroelectricity can be used to make a superconducting switch driven by external electric field \cite{super2022Ferro,2023-MoTe2}. Recent investigations have experimentally revealed that bilayer T$_d$-MoTe$_2$ simultaneously exhibits ferroelectric switching and superconductivity \cite{2023-MoTe2}. Layered transition metal carbides, nitrides and carbon-nitrides named MXenes are an emerging family of 2D materials. Since the synthesis of Ti$_{3}$C$_{2}$T$_{x}$, the progress in the synthesis and development of MXenes crystals with and without the required functional groups have become more mature \cite{Ti3C2Tx, development-MXenes, functionalization-MXenes, functionalization-MXenes-2}, which also promotes the intensity of research on these materials. Surface functionalization can modulate the superconductivity of 2D MXenes \cite{Mo2C3, Nb2C-f, Bekaer2022}.  The fluorinated 2D Nb$_2$N has been reported to have good reversible out-of-plane electrical polarization and metallic properties in the H$_3$ phase \cite{Nb2NF2}. Thus, MXenes have the potential application in low-power-consumption ferroelectric-based superconductor electronic devices. 
    
    The coexistence of superconductivity and nontrivial band topology is highly desired for exploring novel exotic quantum physics \cite{fengyuanping,CaBi2,zhangzhenyu}. A tunable superconductivity and band topology by ferroelectric has immense application potential for superconducting diodes or topological quantum computation driven by external electric field. So far, few 2D ferroelectric superconductivity materials with nontrivial band topology have been predicted or discovered except for ferroelectric-superconductivity heterobilayers, such as IrTe$_{2}$/In$_{2}$Se$_{3}$ \cite{zhangzhenyu}. Theoretically, ferroelectric-superconductivity heterobilayer was predicted to be a useful scheme to achieve the ferroelectric tunable superconductivity and band topological \cite{zhangzhenyu}. The preparation process of heterojunctions is constraines by lattice matching and other issues, making it more attractive to search for intrinsic ferroelectric topological superconductor. This greatly inspires researchers to explore new kinds of topological superconductors regulated by ferroelectric polarization. It is expected to achieve topological superconductivity controlled by ferroelectric polarization in MXene compounds.
	
	In this work, using first-principles approaches, we systematically studied the superconductivity and band topology of fluorinated Nb$_2$N at different adsorption sites. We demonstrate that H$_3$-Nb$_2$NF$_2$ not only possesses intrinsic superconductivity with $T_{c}$ of 10 K, but also exhibits nontrivial band topology. While, H$_1$-Nb$_2$NF$_2$, which contain the inversion symmetry structure among all dynamically stable structures, has the highest $T_{c}$ of 32 K, surpassing most of superconductors with nontrivial band topological properties. The $T_{c}$ sensitively depends on the ferroelectric polarization phase. Our research is of great significance for the development of ferroelectric and superconducting coupling devices.
	
	\section{Computational details}
	
	The structural and electronic properties of hydrogenated and fluorinated Nb$_2$N are studied based on the density functional theory (DFT) calculations with the projector augmented wave (PAW) method \cite{PAW-1,PAW-2}, as implemented in the VASP package \cite{VASP}. The Perdew-Burke-Ernzerhof generalized gradient approximation (PBE-GGA) \cite{GGA} is employed and the electron-ion interaction is described by using the PAW method. The Fermi surfaces (FS) is broadened by the Gaussian smearing method with a width of 0.05 eV. All the geometries are relaxed, and convergence tolerances of force and energy are set to 0.01 eV and 10$^{-6}$ eV, respectively. A vacuum separation is set to be more than 20 Å. to prevent any interactions between two neighboring monolayers. A 12 $\times$ 12 $\times$ 1 Monkhorst-Pack $k$-point mesh is used to sample the 2D Brillouin zone (BZ). The data processing of energy band,orbital-projected electron band structures, density of state and orbital-projected density of state are carried out by QVASP \cite{qvasp} and VASPKIT \cite{VASPKIT} software packages.
	
	To investigate the phonon dispersion and EPC, the density functional perturbation theory (DFPT) \cite{DFPT} calculations are performed with the Quantum ESPRESSO (QE) package \cite{QE}. The kinetic energy cutoffs of 80 and 800 Ry are chosen for the wave functions and the charge densities, respectively. The Methfessel-Paxton smearing width of 0.02 Ry is used. The BZ $k$-point \cite{KPOINTS} grids of 48 $\times$ 48 $\times$ 1 and 24 $\times$ 24 $\times$ 1 is adopted for the dense and sparse self-consistent electron-density calculations, respectively. The dynamic matrix and EPC matrix elements are calculated on 12 $\times$ 12 $\times$ 1 $q$-point meshes. The FS colored as a function of an arbitrary scalar quantity in this work are drawn by using the FERMISURFER program \cite{FS}. Surface states were calculated by the WANNIERTOOLS package \cite{WannierTools}, with a basis set relying on maximally localized Wannier functions (MLWFs) \cite{MLwannier-functions} from the VASP2WANNIER90 interfaces \cite{vasp2wannier}.
	
	The relevant calculation formulas and calculation details of superconductivity are shown in supplementary materials (SM).

\section{Results and discussion}

\subsection{Structure and stability}

	\begin{figure}
		\centering
		\includegraphics[width=1\linewidth]{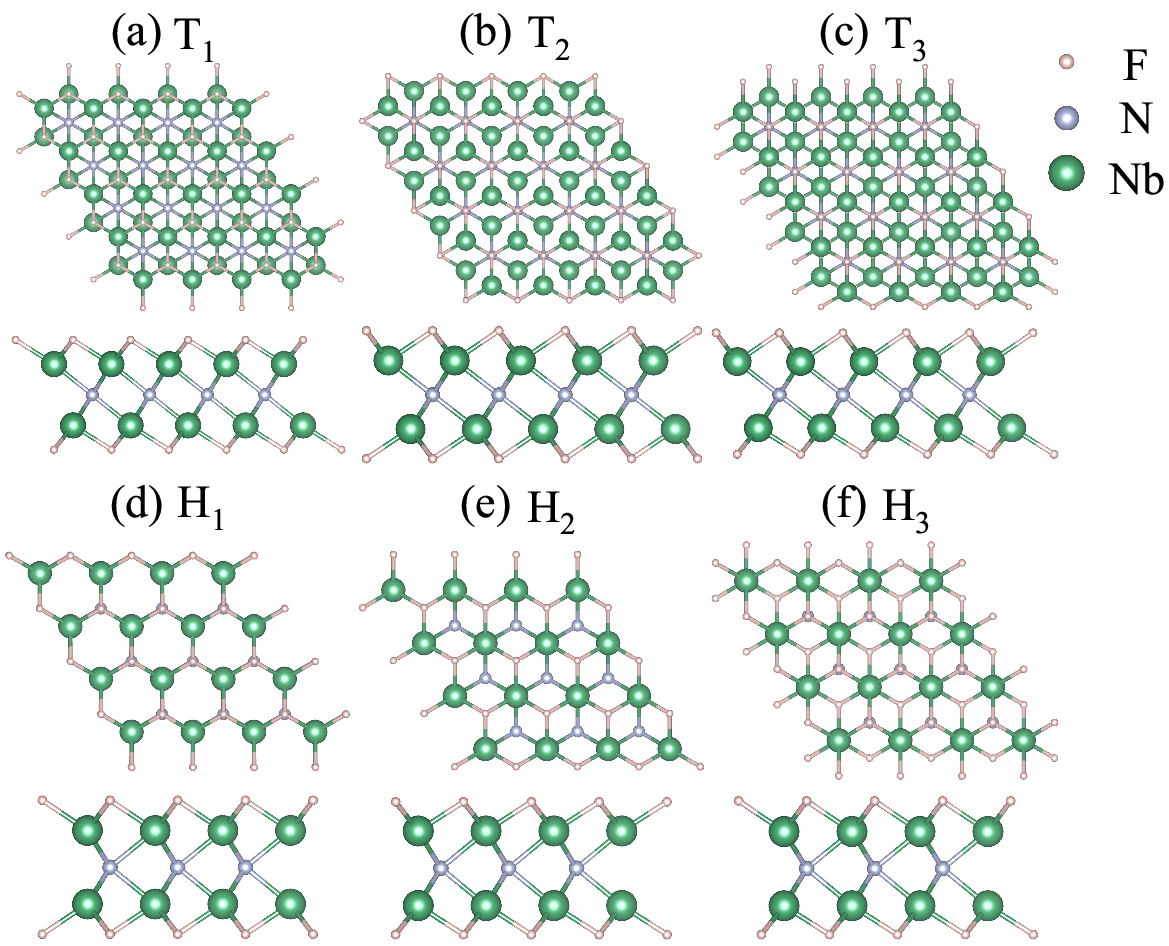}
		\caption{Top (upper panel) and side (lower panel) views of Nb$_{2}$NF$_{2}$ in (a) T$_1$ (b) T$_2$ (c) T$_3$ (d) H$_1$ (e) H$_2$ (f) H$_3$ phases.}
		\label{fig:structure}
	\end{figure}

	In our calculations, we consider all the possible fluorine atom adsorption positions of the 1T- and 2H-Nb$_{2}$N. The stable structures after fluorination of the 1T and 2H phases are named as T$_1$, T$_2$, T$_3$, H$_1$, H$_2$, and H$_3$, respectively, as shown in Fig. 1. The possible adsorption positions are consistent with the previous theoretical results \cite{MXenes-H2, Nb2NF2}, which studied the hydrogenated and fluorinated of MXene. For the T$_1$ phase, the F atoms vertically align with Nb atoms on the other side of the stack. As shown in Fig. 1(b), both F atoms align with the N atom to form the T$_2$ phase. Finally, one F atom aligns with the Nb atom, while the other F atom aligns with the N atom to form the T$_3$ phase. Similarly, the three corresponding H-type configurations are referred to as H$_1$, H$_2$, and H$_3$ structures, as shown in Figs. 1(d-f). Obviously, we can find that T$_3$ and H$_3$ phases break the spatial inversion symmetry. Therefore, they have intrinsic spontaneous ferroelectric polarization. TABLE \uppercase\expandafter{\romannumeral1} contains the lattice constants $a$ and bond lengths of Nb-N, and Nb-F for the corresponding before and after fluorinated stable structures. After fluorinated, the lattice constants and the corresponding N-Nb bond lengths have small change comparing with the ones in Nb$_{2}$N. This very limited perturbation of the lattice due to F is in stark contrast with the observation of giant in-plane lattice expansion of Ti$_2$C with Te functionalization (over than 18\%) \cite{2020science}. The relaxed Nb-F bond lengths is about 2.2 \AA{} in almost all the cases.
		 
	\begin{table}
		\caption{Calculated lattice constant $a$ (\AA), formation energy $E$$_{form}$ (eV/atom) and bond lengths (\AA) of N-Nb and Nb-F.}
		\begin{tabular}{l c c c c}
			\hline
			\hline
			   	& \emph{a} & $E$$_{form}$ & $l$$_{N-Nb}$ &$l$$_{Nb-F}$\\
			\hline
			1T-Nb$_2$N & 3.13 & - & 2.13 & - \\
			2H-Nb$_2$N & 2.90 & - & 2.20 & - \\
			T$_2$-Nb$_2$NF$_2$ & 2.97 & -1.76 & 2.18 & 2.26 \\
			H$_1$-Nb$_2$NF$_2$ & 2.96 & -1.77 & 2.21 & 2.25 \\
			H$_3$-Nb$_2$NF$_2$ & 2.93 & -1.81 & 2.25 & 2.22 2.24 \\
			\hline
			\hline
		\end{tabular}
	\end{table}

	To be a feasible ferroelectric and superconductor, firstly it should be dynamically and thermodynamically stable. To explore the thermo dynamical stability, the formation energy of Nb$_{2}$NF$_{2}$ is calculated by $E_{form}=[E(Nb_{2}NF_{2})-(E(Nb_{2})+\frac{1}{2}E(N_{2})+E(F_{2}))]/5$, where $E(Nb_{2}NF_{2})$, $E(Nb_{2})$, $E(N_{2})$ and $E(F_{2})$ are the energies of the Nb$_{2}$NF$_{2}$, Nb$_{2}$ (space group is Im$\overline{3}$m), N$_{2}$ and F$_{2}$, respectively. The value of $E$\textit{{\footnotesize form}} for thermodynamically stabilized Nb$_{2}$NF$_{2}$ are listed in TABLE \uppercase\expandafter{\romannumeral1}, the negative value proves their thermodynamical stability. For 2D materials, the requisite for a dynamic stable structure is that the phonon dispersion has no imaginary frequency. Then, the dynamic and thermodynamic stabilized Nb$_{2}$NF$_{2}$ are listed in TABLE \uppercase\expandafter{\romannumeral1}, since the absence of obvious imaginary frequencies in the phonon dispersion will be shown later.

\subsection{Electronic structure and ferroelectric polarization properties}
		
	\begin{figure*}
		\centering
		\includegraphics[width=1\linewidth]{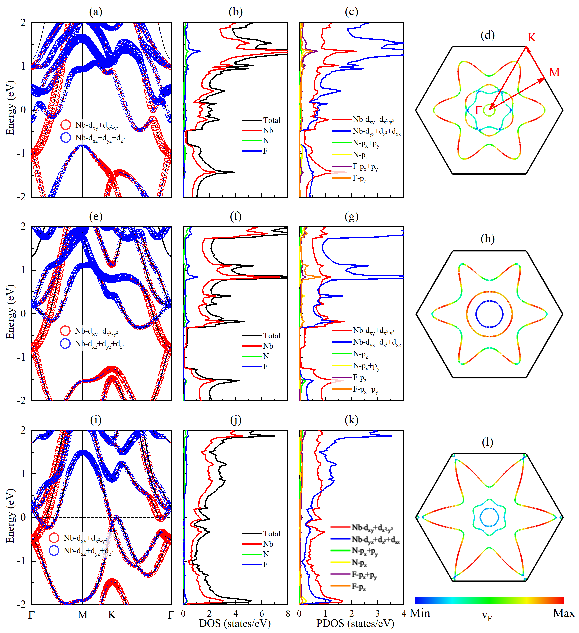}
		\caption{(a) Orbital-projected electronic band structure of Nb along high-symmetry line $\Gamma$-M-K-$\Gamma$, (b) the total DOS and the projected DOS of Nb, N and F, (c) the orbital-partial DOS, and (d) the FS of T$_2$-Nb$_2$NF$_2$. The Fermi level indicated by the dotted line is set to 0 eV, and the color in the FS are proportional to the magnitude of the Fermi velocity v$_F$. (e-h) and (i-l) are similar to (a-d), but for H$_1$- and H$_3$-Nb$_2$NF$_2$, respectively.}
		\label{fig:electric}
	\end{figure*}
	
	Figures 2(a, e, i) show the orbital-projected electron band structures (PBANDs) of T$_2$-, H$_1$-, and H$_3$-Nb$_2$NF$_2$ along high-symmetry line $\Gamma$-M-K-$\Gamma$, respectively. Obviously, all of these structures are metal. Figures 2(b, f, j) and 2(c, g, k) show the total density of states (DOS) and the DOS of Nb, N and F element, and the partial DOS for T$_2$-, H$_1$-, and H$_3$-Nb$_2$NF$_2$, respectively. The DOS of T$_2$-, H$_1$-, and H$_3$-Nb$_2$NF$_2$ at the Fermi level is mainly contributed by Nb-$d$$_{xy}$+$d$$_{{x^2}-{y^2}}$ and Nb-$d$$_{xz}$+$d$$_{yz}$+$d$$_{z^2}$ orbitals, and the contributions of other elements at the Fermi level can be neglected. From the partial DOS, it can be seen that the contributions of in-plane Nb-\textit{d} orbitals are larger than the our-of-plane ones to the band at the Fermi level for both H$_1$- and H$_3$-Nb$_2$NF$_2$, while it is mainly contributed by the out-of-plane Nb-\textit{d} orbitals for the T$_3$-Nb$_2$NF$_2$. These results indicate that there is a strong hybridization between Nb with other elements, which exhibit $sp^{\footnotesize 3}$ hybridization resulting in the metallic $\sigma$ electrons. More interestingly, both the in-plane and out-of-plane of Nb-\textit{d} orbitals in H$_1$-Nb$_2$NF$_2$ are all larger than the ones in H$_3$-Nb$_2$NF$_2$ at the Fermi level because of the differences in spatial inversion symmetry, which may further affect the electron-phonon coupling (EPC). Figures 2(d, h, l) are the FS of T$_2$-, H$_1$-, H$_3$-Nb$_2$NF$_2$, respectively. We can see that there is a circular electron pocket around the $\Gamma$ point of T$_2$-Nb$_2$NF$_2$, and around the $\Gamma$ point, H$_1$-Nb$_2$NF$_2$ is composed of an electron pocket and a hole pocket, with the hole pocket closer to the $\Gamma$ point. The H$_3$-Nb$_2$NF$_2$ exhibits spontaneous electrical polarization because of the breaking of spatial inversion symmetry along the out-of-plane direction. Thus, the H$_3$-Nb$_2$NF$_2$ provides a new platform to study the interplay between symmetry, ferroelectricity and superconductivity. In the subseguent discussion of this work, we will only focus on the possible changed in superconductivity and topological properties duing the ferroelectric polarization reversal of H$_3$ phase.
			
	\begin{figure*}
		\centering
		\includegraphics[width=1\linewidth]{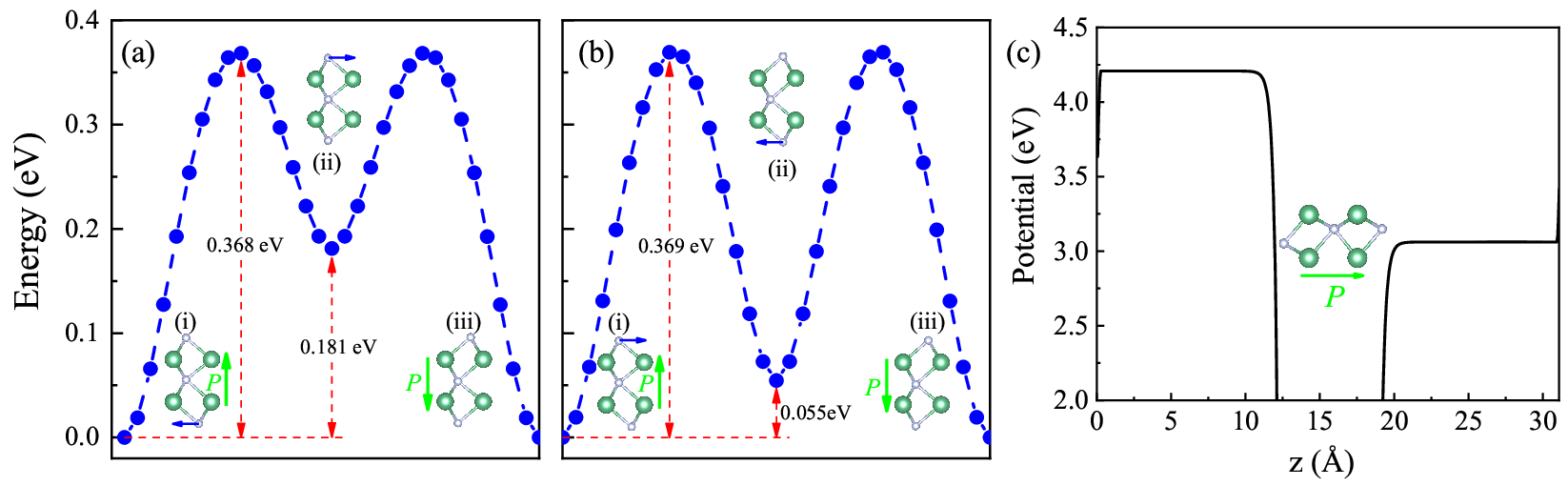}
		\caption{NEB calculations for polarization direction reversal in H$_3$-Nb$_2$NF$_2$, considering (a) the H$_1$ and (b) the H$_2$ non-polar Nb$_2$NF$_2$ as intermediate state. The polarization direction is indicated by the green arrow accompanied by the symbol P. (c) The averaged electrostatic potential of H$_3$-Nb$_2$NF$_2$ in the out-of-plane direction.}
		\label{fig:neb}
	\end{figure*}
	
	Then, we further demonstrate ferroelectricity in H$_3$-Nb$_2$NF$_2$ by investigating the in-plane averaged electrostatic potential in the out-of-plane direction, the possible kinetic pathways and corresponding activation energy barrier of the polarization reversal between different polarized states. TABLE \uppercase\expandafter{\romannumeral2} lists the polar parameters including electric dipole moment $p$, volume $V$, lattice vector $c$, area $S$, and polarization intensity $P$. For H$_3$-Nb$_2$NF$_2$, the polarization intensity is 1.01$\times$10$^{-11}$ C$\cdot$m$^{-1}$, which is comparable with the value of monolayer As or Sb \cite{monolayer-As}.  External electric field can change the polarization direction of ferroelectric materials that can be reversed. Here, we use the nudged elastic band (NEB) method to study the two kinetic pathways and corresponding activation energies for reversing the polarization direction of H$_3$-Nb$_2$NF$_2$. In the process of polarization reversal, the structure is transformed into a intermediate state, e.g. H$_1$ or H$_2$ phases, which are non-polar, symmetric and stable during the transformation. Figures 3(a, b) show the kinetic paths with H$_1$- and H$_2$-Nb$_2$NF$_2$ as intermediate states, respectively. It can be seen that the difference between the potential barriers of these two kinetic paths is only 0.001 eV, and the activation energy barrier of H$_1$ phase is lower than the one of H$_2$ intermediate states. Therefore, the probability of H$_1$ phase as the intermediate state of the kinetic path of H$_3$-Nb$_2$NF$_2$ polarization direction inversion is greater, which is consistent with the previous work by Du $et, al$ \cite{Nb2NF2}. Figure 3(c) illustrates the electrostatic potential of H$_3$-Nb$_2$NF$_2$ with respect to the out-of-plane direction and it shows that the vacuum levels at either end of the structure differ by 1.1 eV. The difference between vacuum levels and its direction (high to low) demonstrates that the ferroelectric polarization is greater than the self-depolarization field generated by the mobile carriers \cite{Nb2NF2}.
	  
	  \begin{table}
	  	\caption{The polarization parameters, including electric dipole moment $p$ (e$\cdot$\AA), volume $V$ (\AA$^3$), lattice vector $c$ (\AA), area $S$ (\AA$^2$), and polarization intensity $P$ (C$\cdot$m$^{-1}$) for ferroelectric H$_3$-Nb$_2$NF$_2$.}
	  	\centering
	  	\begin{tabular}{l c c c c c}
	  		\hline
	  		\hline
	  		       & $p$ & $V$ & $c$ & $S$ & $P$ \\
	  		\hline
	  		H$_3$-Nb$_2$NF$_2$	& 0.047 & 231.64 & 31.106 & 7.447 & 1.01$\times$10$^{-11}$ \\
	  		\hline
	  		\hline
	  	\end{tabular}
	  \end{table}	
	  
\subsection{Electronic-phonon coupling and possible superconductivity}
			
	\begin{figure*}
		\centering
		\includegraphics[width=1\linewidth]{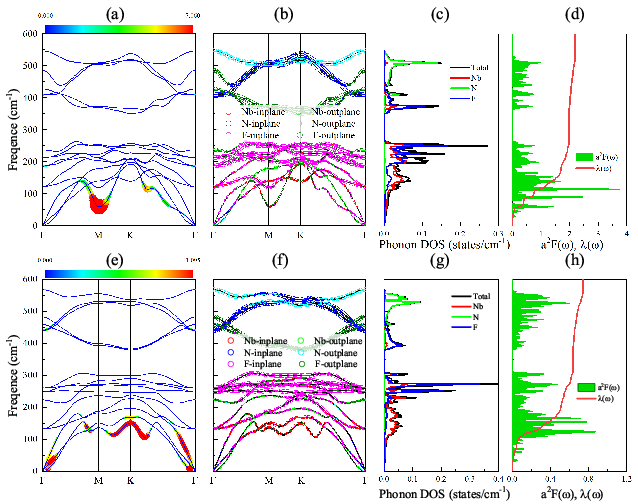}
		\caption{(a) Phonon dispersion weighted by the magnitude of EPC $\lambda$$_{q\nu}$ (the magnitude of $\lambda$$_{q\nu}$ is displayed with different scale in different figures), (b) phonon dispersion weighted by the in-plane and out-plane vibration modes of Nb, N and F elements, (c) total and atom-projected phonon DOS, (d) Eliashberg spectral function a$^2$$F$($\omega$) and cumulative frequency-dependent EPC function $\lambda$($\omega$) of H$_1$-Nb$_2$NF$_2$. (e-h) are similar to (a-d) but for H$_3$-Nb$_2$NF$_2$}
		\label{fig:phonon}
	\end{figure*}

	Since T$_2$-, H$_1$- and H$_3$-Nb$_2$NF$_2$ are all metal, we further explore the superconductivity of them. The phonon dispersion weighted by the magnitude of EPC $\lambda$$_{q\nu}$, the in-plane and out-of-plane vibration modes contributed by different elements, the corresponding total and atom-projected phonon DOS, Eliashberg spectral function a$^2$$F$($\omega$) and cumulative frequency-dependent EPC function $\lambda$($\omega$) are shown in Fig. 4. The properties of T$_2$-Nb$_2$NF$_2$ are included in the SM. For Nb$_2$NF$_2$, there are 5 atoms in each primitive cell, generating 15 phonon branches, with the frequency range of phonon dispersion up to 600 cm$^{-1}$. The phonon dispersion weighted by the in-plane and out-plane vibration modes of Nb, N and F elements of H$_1$- and H$_3$-Nb$_2$NF$_2$ are shown in Figs. 4(b, f), where the vibration modes can be clearly divided into two parts. The low-frequency range (0 $<$ $\omega$ $<$ 320 cm$^{-1}$) mainly originates from the vibration modes of Nb and F elements, while the contribution of N element is relatively small. And, the high-frequency range (320 $<$  $\omega$ $<$ 600 cm$^{-1}$) mainly contributed by vibration modes of N and F elements. Combined with the phonon density of state (PhDOS) analysis, it can be seen that the contribution of elements to phonon dispersion is closely related to the atomic mass for both cases, as shown in Figs. 4(c, g). The contribution of low-frequency part to phonon dispersion mainly comes from Nb element with larger atomic mass, the contribution of mid-frequency part comes from the F element, and the contribution of high-frequency part comes from the N element. Obviously, the positions of peaks in the Eliashberg spectral function are highly consistent with those of the PhDOS. While, the Eliashberg spectral function and cumulative frequency-dependent EPC $\lambda$($\omega$) are different for H$_1$- and H$_3$-Nb$_2$NF$_2$. For H$_1$-Nb$_2$NF$_2$, the frequency-dependent EPC function $\lambda$($\omega$) indicate that its EPC that of from the low-frequency acoustic branch is stronger then H$_3$-Nb$_2$NF$_2$, which can also be seen from Figs. 4(a) and 4(e). It results in H$_1$-Nb$_2$NF$_2$ having a higher $T_{c}$ than H$_3$-Nb$_2$NF$_2$.
	
	\begin{figure}
		\centering
		\includegraphics[width=1\linewidth]{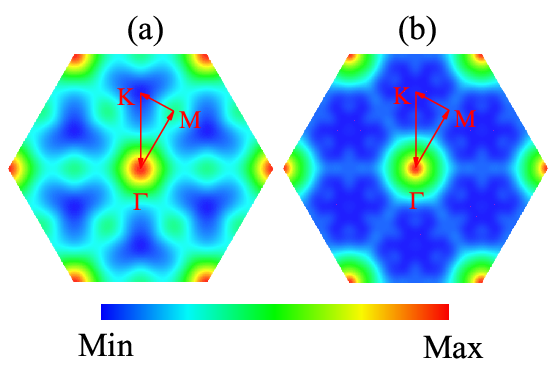}
		\caption{FS nesting of (a) H$_1$- and (b) H$_3$-Nb$_2$NF$_2$.}
		\label{fig:nesting}
	\end{figure}
		
	\begin{table}
		\caption{Logarithmic averaged phonon frequency $\omega_{log}$ (K), electronic DOS at the Fermi level $N$($E_{F}$) (eV$^{-1}$), total EPC constant $\lambda$, strong-coupling correction factor $f_1$, shape correction factor $f_2$ and estimated $T_{c}$ (K) for 2D Nb$_2$NF$_2$.}
		\centering
		\begin{tabular}{l c c c c c c}
			\hline
			\hline
			       & $\omega_{log}$ & $N$($E_{F}$) & $\lambda$ & $f_1$ & $f_2$ & $T_{c}$ \\
			\hline
			H$_1$-Nb$_2$NF$_2$ & 180.81 & 2.03 & 2.19 & 1.15 & 1.01 & 32.1  \\
			H$_3$-Nb$_2$NF$_2$	& 245.57 & 1.81 & 0.74 & 1.03 & 1.00 & 10.1 \\
			\hline
			\hline
		\end{tabular}
	\end{table} 
	
	Combined with the phonon dispersion weighted by the magnitude of EPC $\lambda$$_{q\nu}$, the low frequency phonon branches have been softened in H$_1$-Nb$_2$NF$_2$ as compared with the stabilized H$_3$-Nb$_2$NF$_2$. The calculated Eliashberg spectral function a$^2$$F$($\omega$) and cumulative frequency-dependent EPC function $\lambda$($\omega$) of H$_1$- and H$_3$-Nb$_2$NF$_2$ are shown in Figs. 4(d) and 4(h), respectively. The $\lambda$ of H$_1$-Nb$_2$NF$_2$ with high symmetry is 2.91, leading to $T_{c}$ of 32.1 K. While, the $\lambda$ of H$_3$-Nb$_2$NF$_2$ with the breaking of spatial inversion symmetry is 0.741, and the corresponding $T_{c}$ is 10.1 K. By analyzing the FS, the total DOS at the Fermi level, phonon dispersion, and EPC strength of Nb$_2$NF$_2$ with different spatial inversion symmetry, there are four important aspects related to the enhancement of $T_{c}$ in H$_1$-Nb$_2$NF$_2$. Firstly, the spatial inversion symmetry will have a significant impact on the charge distribution of the system, especially around the Fermi level. As well known, the ferroelectric polarization originated from the breaking spatial inversion symmetry. Then, the DOS at the Fermi level and the morphology of FS are different for the polarized H$_3$-Nb$_2$NF$_2$ and the unpolarized H$_1$-Nb$_2$NF$_2$.  Secondly, as is known to all, the increasing DOS $N$($E_{F}$) at the Fermi level will be conducive to the EPC $\lambda$. As summarized in TABLE \uppercase\expandafter{\romannumeral3}, the $N$($E_{F}$) of H$_1$-Nb$_2$NF$_2$ is enhanced by more than 12.3 $\%$ from that of H$_3$-Nb$_2$NF$_2$. The increased $N$($E_{F}$) may further give rise to an enhancement of FS nesting. Thirdly, the FS nesting is beneficial to realize the pairing of electron for the superconductivity. Fourthly, the phonon branches contributed by vibrations of Nb is significant changed because of the different spatial inversion symmetry between H$_1$-Nb$_2$NF$_2$ and H$_3$-Nb$_2$NF$_2$, resulting in an enhancement of $\lambda$ according to Eq.(4) of SM. 

	To further investigate the EPC of H$_1$- and H$_3$-Nb$_2$NF$_2$, in Figs. 5(a) and 5(b), the imaginary parts of the bare electronic susceptibility $\chi$$''$(q) are shown, which can directly evaluate the FS nesting in the low-frequency limit. As for $\chi$$''$(q), it does carry out entire FS nesting into itself near the $\Gamma$ point, which has no actual physical meaning. For H$_1$-Nb$_2$NF$_2$ another large value of FS nesting appears near the M point, consistent with the large $\lambda$$_{q\nu}$ around M. To be more clear, in Figs. 4(a) and 4(b), about 60 cm$^{-1}$, there are softened low-energy phonon branches along high-symmetry line $\Gamma$-M-K-$\Gamma$, especially at the M point. The phonons of these positions show large $\lambda$$_{q\nu}$ in Fig. 4(a). Meanwhile, the imaginary part of the electronic susceptibility, reflecting the nesting effect, shows similar distribution. Therefore, we infer that the nesting effect accounts for the softened phonons in H$_1$-Nb$_2$NF$_2$ and the presence of large EPC $\lambda$$_{q\nu}$. This phenomenon has also been proposed in transition metal chalcogenides \cite{1T-VSe2,2012-VSe2,1T-TaS2}. Thus, we can conclude that strong FS nesting is responsible for the strong EPC in H$_1$-Nb$_2$NF$_2$. 
	
\subsection{Topological properties}

	Then, we investigate the band topology in the 2D Nb$_2$NF$_2$. The electronic band structures with and without spin-orbit coupling (SOC) for H$_1$- and H$_3$-Nb$_2$NF$_2$ along high-symmetry directions are shown in Figs. 6(a) and 6(b), respectively. The bands near the Fermi energy are considered. Both for H$_1$- and H$_3$-Nb$_2$NF$_2$, the band overlap occurs in the $\Gamma$ point below Fermi energy as indicated by a green circle. When SOC is induced, although there is no global gap, a gap-opening occurs at each point in the whole Brillouin zone for both cases. The gap-opening with SOC may be accompanied by a topological phase transition. Then, the topological invariant $Z_2$ is computed based on the Wannier charge-centers(WCC) method \cite{Z2-invariant}, which is suitable for the structures with and without spatial inversion symmetry. For H$_1$- and H$_3$-Nb$_2$NF$_2$, any arbitrary horizontal reference line crosses an odd number of times of WCC, which means a topologically nontrivial state ($Z_2$=1). Figures 6(c) and 6(d) show the edge states of H$_1$- and H$_3$-Nb$_2$NF$_2$ with a semi-infinite slab. The presence of nontrivial topological edge states within the bulk gap can further support the nontrivial band topology. Thus, H$_3$-Nb$_2$NF$_2$ provides a realistic platform for experimental investigations into topological superconductors with tunable ferroelectric properties.
				
	\begin{figure}
		\centering
		\includegraphics[width=1\linewidth]{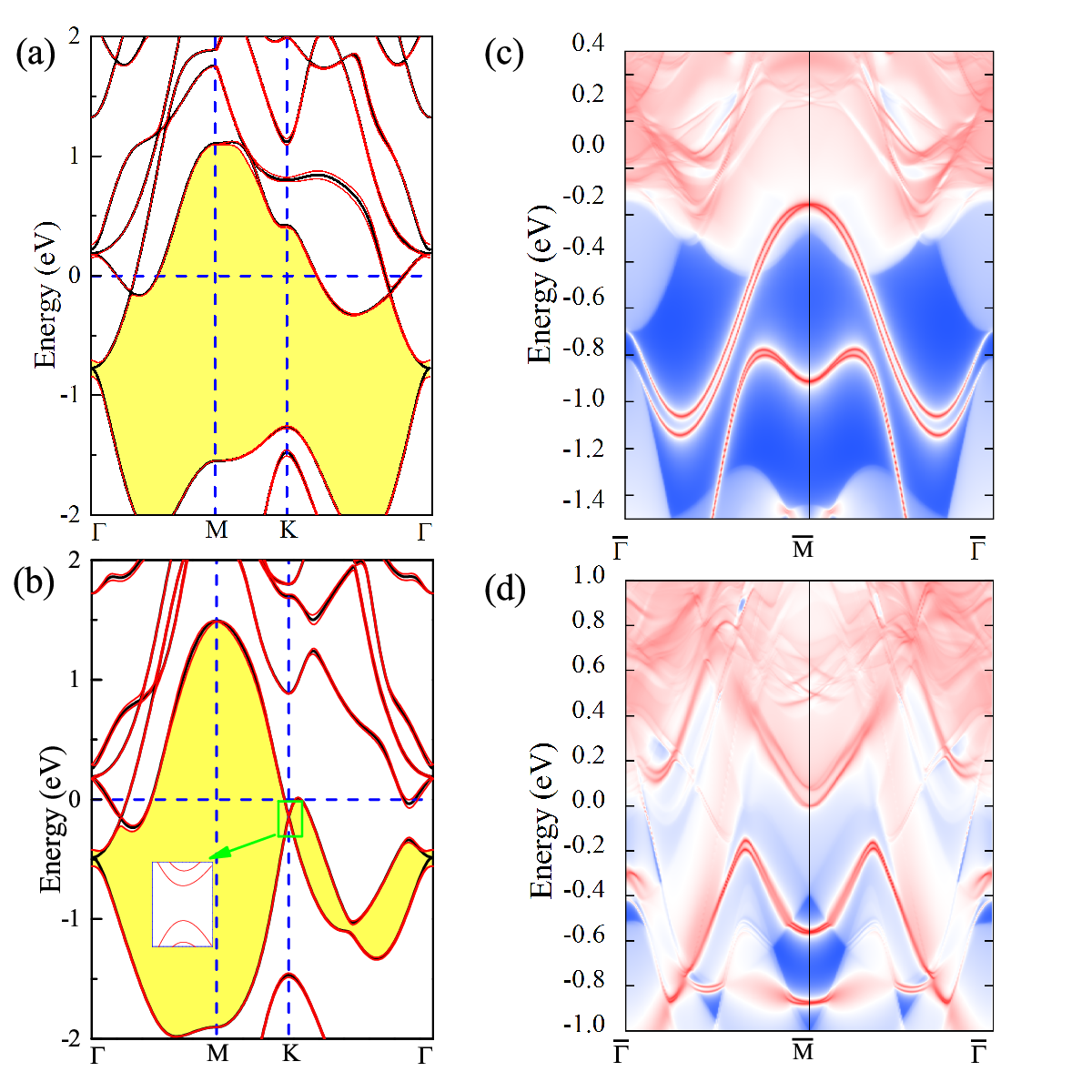}
		\caption{Band structures for (a) H$_1$- and (b) H$_3$-Nb$_2$NF$_2$ without (black) and with (red) the SOC. The full-gap near the Fermi level for both structures with SOC are decorated by yellow. Edge states  of the corresponding semi-infinite slab for (c) H$_1$- and (d) H$_3$-Nb$_2$NF$_2$.}
		\label{fig:topology}
	\end{figure}
	
\section{Discussion and conclusion}
	
	Recent investigations have revealed that the external electric field can be a tuning knob for superconductivity in 2D heterojunctions such as bilayer T$_d$-MoTe$_2$ \cite{2023-MoTe2} and twisted graphene \cite{Park2021}. Some of these systems show ferroelectricity properties driven by external electric field. More interestingly, the ferroelectricity of 2D In$_2$Se$_3$/IrTe$_2$ heterojunctions can not only modulate superconductivity, but also switch its topological electronic structure between trivial and nontrivial \cite{zhangzhenyu}. However, the preparation of high-quality heterojunctions is very difficult due to the lattice matching and interface issues. It is very important to find 2D monolayer intrinsic topological superconductors' candidates coupled with ferroelectric polarization.
	
	In this work, we have studied the electronic properties and superconductivity of the fluorinated structures of Nb-based MXenes using first-principles calculations and further investigated the ferroelectric properties of the spatial inversion symmetry breaking structure. The following conclusions can be obtained: (i) Surface fluorination can induce superconductivity in Nb$_2$N. (ii) High symmetry is benificial for a structure to show high-$T_{c}$ superconductivity. (iii) There is potential ferroelectricity in the asymmetric phase where spatial inversion symmetry is broken. Our research has enriched 2D superconducting materials with nontrivial band topology and ferroelectric properties, and provided a theoretical basis for the preparation of devices switching between superconducting and ferroelectric states with external electric field.
		
	\begin{acknowledgements}		
		This work is supported by the National Natural Science Foundation of China (Grant Nos. 12074213, 11574108, 12104253), the Major Basic Program of Natural Science Foundation of Shandong Province (Grant No. ZR2021ZD01), the Natural Science Foundation of Shandong Provincial (Grant No. ZR2023MA082) and the Project of Introduction and Cultivation for Young Innovative Talents in Colleges and Universities of Shandong Province.		
	\end{acknowledgements}	
	
	\vspace{0.5cm}$^*$E-mail: j$\_$n2013@126.com, hylu@qfnu.edu.cn
	
	\bibliographystyle{apsrev4-1}
	\bibliography{References}
	
\end{document}